\documentclass[10pt,twocolumn]{article}
\usepackage[T1]{fontenc}
\usepackage{microtype}
\usepackage[margin=0.75in]{geometry}
\usepackage{amsmath,amssymb,amsthm,graphicx,booktabs,url}
\usepackage[colorlinks=true,citecolor=blue,urlcolor=blue]{hyperref}
\newtheorem{proposition}{Proposition}
\newcommand{\E}{\mathbb E}
\newcommand{\Prob}{\mathbb P}
\usepackage{abstract}

\title{When Does Interaction Help?\\Toward a Reynolds Number for Collective Intelligence}

\author{Xuening Wu$^{1}$\thanks{Correspondence: \href{mailto:xuening.wu@pfizer.com}{\nolinkurl{xuening.wu@pfizer.com}}}, Haode Yang$^{1}$, Yanlan Kang$^{2}$, Shenqin Yin$^{2}$\\[0.5em]
\small $^{1}$Pfizer Inc.\quad $^{2}$Fudan University}
\date{Working manuscript, November 14, 2026}
\begin{document}
\maketitle

\begin{abstract}
\noindent
Multi-agent systems combine distributed information through interaction, which can improve performance but also spread errors or suppress useful differences. Predicting when interaction helps is central to understanding and designing collective intelligence.
We investigate whether a few dimensionless parameters can predict gains over explicitly defined non-communicating baselines across group sizes, network structures, and tasks, toward a Reynolds-number-like description of collective intelligence. Using distributed classification tasks, stochastic tracking networks, and neural cellular automata, we combine analytical and empirical evidence to examine information sufficiency, dynamical applicability, and finite-sample prediction.
We construct inference systems with bounded random variables and matching specified joint moments up to total degree three, yet opposite interaction gains under a fixed readout. These moments alone cannot generally determine the gain sign. Tracking networks likewise exhibit gain reversals despite matched candidate parameters when task power occupies different spectral modes. Under concentrated task spectra and aligned observation noise, however, we derive a low-dimensional similarity relation with an explicit zero-delay error bound and test it across network sizes and topologies under frozen protocols. The relation does not directly apply to the neural cellular systems, where a separate moment-based predictor underperforms direct gain measurement at matched calibration sample budgets.
These results establish conditions and limits for low-dimensional gain prediction, providing a restricted similarity law while leaving a general Reynolds-number-like description of collective intelligence open.
\end{abstract}

\section{Introduction}\label{sec:intro}

Local interaction can combine complementary evidence, suppress noise, or spread errors. Understanding when it improves collective performance requires more than counting agents or measuring connectivity. A compact predictive description would allow comparisons across network sizes, topologies, and task timescales without evaluating every configuration from scratch.

The search for a ``Reynolds number for collective intelligence'' expresses this ambition: independently meaningful dimensionless quantities might organize collective-performance curves across systems. The analogy concerns dynamical similarity, not a reduction in state dimension. In fluid mechanics, Reynolds-number similarity requires corresponding dimensionless geometry, initial and boundary conditions, and governing assumptions; equal Reynolds numbers alone do not imply equivalent behavior across arbitrary flows \cite{ceniceros2011}. Likewise, we seek classes of collective systems whose performance can be organized by a few dimensionless parameters under explicit structural and task conditions.

Dimensional consistency alone does not guarantee predictive value. A compact description must retain task-relevant information, its prediction rule must apply to the system under study, and its parameters must be estimated accurately enough to support useful prediction. These requirements motivate our three questions:
information sufficiency, dynamical applicability, and finite-sample prediction. Success at one level does not ensure success at the next.

\begin{figure*}[!ht]
    \centering
    \includegraphics[width=0.8\textwidth]{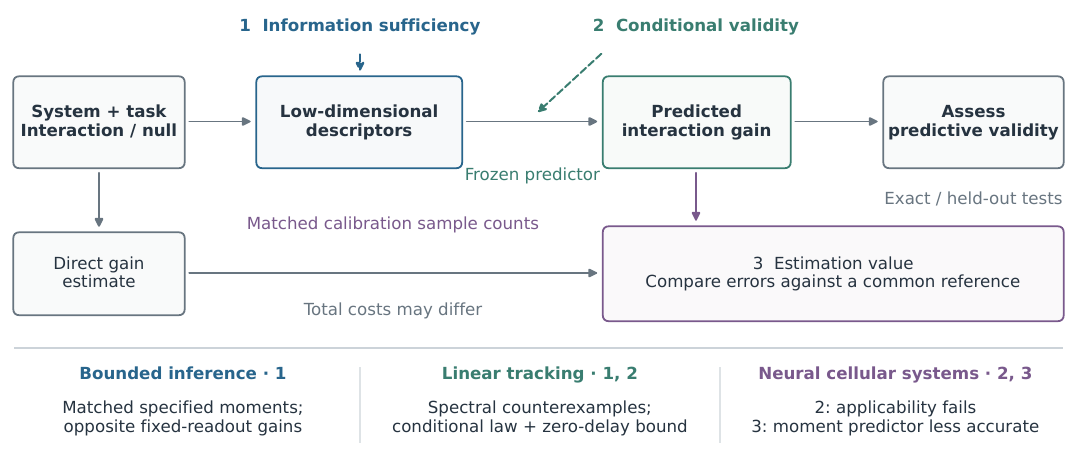}
    \caption{Framework for assessing interaction-gain prediction:
    information sufficiency, conditional validity, and estimation
    value. The model classes provide complementary evidence.
    Direct estimation serves as a comparator at matched calibration
    sample counts. Solid arrows indicate prediction and comparison
    paths; dashed arrows indicate checks.}
    \label{fig:overview}
\end{figure*}

Prior work on Renormalized Cellular Intelligence (RCI) \cite{wu2026} studies preservation of task-relative capability under model reduction. Here we ask when compact descriptors can predict interaction gains across conditions. Probe~3 in that study already demonstrates distribution-dependent gains at identical response descriptors and derives a lower bound on descriptor-only prediction error. We strengthen this obstruction with bounded systems whose specified joint raw moments through degree three match exactly, yet whose fixed-readout gains have opposite signs. A separate comparison establishes a Bayes-optimal gain separation despite the same specified moments.

The model classes provide complementary evidence. The bounded inference construction isolates information lost by specified descriptors. Linear tracking networks permit exact analysis of task spectra, timescales, and conditional similarity. Frozen neural cellular automata test prediction at matched calibration budgets and the applicability of the tracking assumptions. These models do not constitute a validated coarse-graining chain.

Figure~\ref{fig:overview} summarizes the three questions and the complementary roles of the model classes.

Our contributions are:
\begin{enumerate}
\item \textbf{Explicit identifiability obstructions.}
We construct bounded systems with identical specified joint moments through degree three but opposite fixed-readout gains, separately establish a Bayes-gain separation, and exhibit tracking systems with matched candidate descriptors and opposite gains.

\item \textbf{Conditional gain similarity and its boundaries.}
For a restricted tracking class, we specialize standard spectral and approximation tools to derive a low-dimensional gain curve with an explicit bandwidth error bound. Tests under frozen protocols include controls that violate the spectral assumptions and distinguish task-response similarity from delay-dependent internal stability.

\item \textbf{Finite-budget and applicability tests.}
Frozen transfer tests and comparisons with paired direct gain measurement distinguish population sufficiency, cross-task prediction, and estimation efficiency. Failures on the native cellular systems delimit the applicability of the tracking law and the usefulness of a separate moment-based predictor.
\end{enumerate}

Our objective is a task-conditioned predictive law with testable assumptions. A universal control number, evidence of a phase transition in intelligence, and an effective reduction connecting the three model classes remain beyond the present results. We introduce the models and predictors in Section~\ref{sec:methods}, organize their evaluation around three questions in Section~\ref{sec:experiments}, and discuss the implications in Section~\ref{sec:discussion}.

\section{Related Work}\label{sec:related}
\paragraph{Collective scalability.}
Hamann and Reina \cite{hamann2021} connect microscopic interactions and coordination overheads to collective scalability. Their work establishes a precedent for compact performance laws grounded in local mechanisms. We focus on the signed gain of a specified interaction relative to a declared null and ask whether proposed descriptors can distinguish systems whose gains differ.

\paragraph{Spectral performance and estimation.}
Xiao et al.\ \cite{xiao2007} analyze steady-state mean-square deviation in noisy consensus. Perraudin and Vandergheynst \cite{perraudin2017} use graph-stationary signal models for Wiener-type estimation, while Routtenberg \cite{routtenberg2026} develops shared spectral responses across network sizes and unseen graphs. Delay-dependent spectral performance is studied by Ghaedsharaf et al.\ \cite{ghaedsharaf2019}. Our conditional similarity result builds on this established spectral perspective. Its specific object is a signed task-gain curve, with matched-descriptor counterexamples, an explicit task-band approximation, and a separate check of all-mode stability.

\paragraph{Statistical sufficiency and information restrictions.}
The elliptical sign-correlation identity \cite{schmid2006} supplies a restricted sufficient description and serves here as a positive reference. Finite moments do not generally determine arbitrary distributional functionals. Our bounded construction makes this obstruction concrete for a distributed task with fixed information permissions, a declared decoder, and a locally optimal null. It separates decoder-specific gain from the gain achievable by a distribution-aware optimal decoder.

\paragraph{Cellular computation and model reuse.}
Neural cellular automata demonstrate how shared learned local rules can produce collective computation \cite{mordvintsev2020}. We reuse the inference probes, frozen co-reachability coefficients, and five localization checkpoints from the RCI study \cite{wu2026}, together with its task-relative interaction/null comparison. The localization architecture includes absolute coordinates and a fixed global mean readout. Reusing these checkpoints permits a controlled test of predictor performance and applicability on the existing models; it does not provide a new independent population of learned systems.

\section{Methods}\label{sec:methods}
We specify the task comparison, the information available to each predictor, and the assumptions under which an analytical description is valid. The shared notation makes comparisons explicit without imposing common dynamics or a common task on the three model classes.

\subsection{Problem formulation and agent interfaces}
An agent receives local observations $o_i^t$, maintains state $z_i^t$, exchanges messages $m_i^t$, and produces a local output $r_i^t$:
\begin{align}
 m_i^t&=\operatorname{Agg}_{j\in N(i)}M_{ji}(z_j^t;\alpha),\nonumber\\
 z_i^{t+1}&=F_i(z_i^t,o_i^t,m_i^t;\alpha),\nonumber\\
 r_i^t&=R_i(z_i^t),\quad \hat y^t=D(r_1^t,\ldots,r_n^t).
\end{align}
The explicit dependence of $F_i$ on intervention $\alpha$ permits interventions that also change self-state mixing. A continuous-time model uses the corresponding stochastic differential equation. This notation does not impose a shared state space, decoder, or task across models.

For a specified task and quality score, define
\begin{equation}
G=\E[Q_I-Q_N],
\end{equation}
where $I$ is the interacting system and $N$ its declared null. Positive $G$ means improvement over that null, not necessarily task adequacy or superiority to an independently optimized system. Classification uses correctness, tracking uses negative node-averaged squared error, and cellular localization uses a normalized distance score. Their numerical gains are not pooled.

\subsection{Models, information access, and baselines}
\paragraph{Distributed inference.}
The transfer experiments reuse the inference graph and role constructions identified in Section~\ref{sec:related}; their task and observation rules are specified in Appendix~\ref{app:transfer}. The exact construction studied here is a separate two-node decision problem: one node observes $(U,W)$ and another observes $V$. It has a fixed terminal decoder and no intrinsic coordination timescale. It is a witness to an information restriction, not a miniature dynamical colony.

\paragraph{Linear tracking.}
On a symmetric connected graph with average-degree-normalized combinatorial Laplacian $L$, the target $x$ has independent Ornstein--Uhlenbeck modes in the eigenbasis $U$ of $L$:
\begin{align}
 dx&=-\omega x\,dt+\sqrt{2\omega}U\operatorname{diag}(\sqrt{s_k})dW,\nonumber\\
 dy_i&=x_i\,dt+\sqrt{2v/a}\,dV_i,\nonumber\\
 dz&=a(dy-z\,dt)-cLz\,dt,\qquad \sum_k s_k=n.
\end{align}
The target and observation-noise Brownian motions are independent. Each node estimates its own target component. The null uses the same observations and local response $a$, with $c=0$. Averaging node losses is an evaluation operation, not a global decoder. The graph-aligned target covariance, common response rate, and linear smoothing strategy are substantive assumptions. The null is not reoptimized for each task.

\paragraph{Neural cellular localization.}
The five reused checkpoints use a $32\times32$ open-boundary lattice, eight state channels, four-neighbor means, and 32 synchronous updates. The static local input contains a noisy masked signal, the mask, and absolute coordinates. With shared neural rule $f_\theta$,
\begin{align}
 h_i^t&=(1-\lambda)z_i^t+\lambda\operatorname{mean}_{j\in N(i)}z_j^t,\nonumber\\
 z_i^{t+1}&=(1-\eta_c)z_i^t+\eta_c f_\theta(h_i^t,o_i),\nonumber\\
 \hat y&=n^{-1}\sum_i (z_{i,1:2}^{32}+1)/2.
\end{align}
Here $\eta_c=0.5$ is a state-update coefficient, not the tracking noise parameter. Targets and sensor noise remain fixed throughout an episode. The quality is $Q=1-\|\hat y-y\|/\sqrt2$. The comparison switches $\lambda=0.5$ to zero while preserving weights, input, and global mean readout. It removes neighbor mixing while changing the self-state coefficient. Thus it measures this architectural intervention, not a pure information-only intervention or superiority over the best independent learner.

\subsection{Prediction rules and information budgets}\label{sec:predictors}
Let $B$ denote the number of target-condition calibration examples. For paired observations indexed by $\ell$, the direct reference is
\begin{equation}
\widehat G_{\rm direct}=B^{-1}\sum_{\ell=1}^{B}(Q_{I\ell}-Q_{N\ell}).
\end{equation}
This estimator uses task labels and both system outputs. Matching $B$ equates target-condition sample counts; it does not equate extra offline training, access to generating parameters, or computation. Table~\ref{tab:predictors} summarizes the comparison objects.

\begin{table*}[t]
\centering\small
\begin{tabular}{p{.16\textwidth}p{.32\textwidth}p{.43\textwidth}}
\toprule
Predictor & Inputs and mapping & Information and fitting\\\midrule
Co-reachability & Structural $(C,H_c,B_c)$ and a fixed polynomial & Graph/propagation quantities; previously fitted coefficients, kept fixed \\
Task reliability & Channel sign reliabilities; $\frac12\operatorname{mean}_i(r_{ai}r_{bi})$ or a fitted polynomial & Calibration component labels; fitted variant uses Gaussian training runs\\
Margin3 & Margin correlations, dispersion and bias; $\beta^\top\phi$ & Both labeled arms; separate ridge fit for each budget, frozen after Gaussian-sum training\\
Elliptical plug-in & Two Pearson correlations in the arcsin formula & Both labeled arms; no fitted gain map\\
Tracking ratio / compact & Reference curve / collapsed nonzero-mode spectrum & Known generating parameters; reference task-class information or six scalars\\
Conditional similarity & $F(R,\eta,b)$ and task-band error bound & Known parameters and aligned-noise assumptions; no label fitting\\
Cellular Gaussian residual & Mean and covariance per residual arm; expected norm difference & Ten estimated parameters from paired labeled outputs; no learned gain map\\
Direct & Paired sample-average quality difference & Same target-condition examples where budgets match\\
\bottomrule
\end{tabular}
\caption{Prediction rules and information permissions. Different rows are evaluated on their respective tasks, not pooled into a single leaderboard.}\label{tab:predictors}
\end{table*}

\paragraph{Structural and task-reliability descriptors.}
For the factorized inference network, $K=\gamma^T P^T$ is the finite-horizon propagation matrix, where $P$ is the propagation matrix, $\gamma$ is the memory factor, and $T$ is the update horizon; the superscript $T$ denotes a power, not a transpose. The parameter $\sigma$ is the readout-noise standard deviation. For binary role vector $b$, set $s_a=K(1-b)$, $s_b=Kb$ and $v_a=K^{\circ2}(1-b)+\sigma^2\mathbf1$, $v_b=K^{\circ2}b+\sigma^2\mathbf1$, where $K^{\circ2}$ is the elementwise square. Define $u_{ai}=s_{ai}/\sqrt{v_{ai}}$, $u_{bi}=s_{bi}/\sqrt{v_{bi}}$, and $w_i=\min(u_{ai},u_{bi})$. Co-reachability uses
\[
 C=\operatorname{mean}_i\frac{w_i}{1+w_i},\quad
 H_c=\operatorname{sd}_i\frac{w_i}{1+w_i},
\]
\[
 B_c=\operatorname{mean}_i\frac{|u_{ai}-u_{bi}|}{1+u_{ai}+u_{bi}}.
\]
The co-reachability predictor applies a frozen linear combination of $(C,C^2,C^3,CH_c,CB_c,CH_cB_c)$, clipped to $[0,1/2]$. The original frozen coefficients and their feature order are given in Appendix~\ref{app:coefficients}.

Task reliabilities are $r_{ai}=\operatorname{mean}_\ell[\operatorname{sign}(o_{ai\ell})\operatorname{sign}(z_{a\ell})]$, and analogously for channel $b$. The analytic factorized reference uses $\frac12\operatorname{mean}_i(r_{ai}r_{bi})$. The fitted task3 variant instead summarizes the mean and standard deviation of $r_{ai}r_{bi}$ and the mean absolute difference of channel reliability magnitudes, applies the same six-term feature pattern, and uses ridge penalty $0.01$ with Gaussian training data. Its outputs in the factorized study are clipped to $[-1/2,1/2]$. It receives component labels unavailable to structural descriptors.

\paragraph{Margin3.}
For task margin $X$ and nodewise interaction/null margins $I_i,N_i$, estimate correlations $\hat\rho_{XI_i},\hat\rho_{XN_i}$ from the same $B$ calibration examples and define
\begin{align}
 g_i&=[\arcsin\hat\rho_{XI_i}-\arcsin\hat\rho_{XN_i}]/\pi,\nonumber\\
 D&=\operatorname{mean}_i g_i,\quad H=\operatorname{sd}_i g_i,\nonumber\\
 J&=\operatorname{mean}_i\left(\frac{|\bar I_i|}{s_{I_i}}-\frac{|\bar N_i|}{s_{N_i}}\right).
\end{align}
Here bars denote calibration-sample means, and $s_{I_i},s_{N_i}$ are calibration-sample standard deviations computed with divisor $B$. Then $\phi=(D,D^2,D^3,DH,DJ,DHJ)^\top$ and
$\widehat G_{\rm margin3}=\operatorname{clip}_{[-1,1]}(\beta^\top\phi)$.
For each budget, $\beta=(\Phi^\top\Phi+0.01I)^{-1}\Phi^\top g_{\rm train}$ is fitted only on Gaussian-sum training runs, then frozen. The rows of $\Phi$ are the training feature vectors $\phi^\top$, and $g_{\rm train}$ contains the corresponding evaluation-sample gains. Here $J$ renames the code's bias descriptor to avoid confusing it with sample budget $B$. Correlations are clipped to $\pm0.999999$; zero-variance correlations and standardized output means are assigned zero according to the fixed implementation. Task margins and arm-specific normalization are detailed in Appendix~\ref{app:transfer}.

\paragraph{Cellular residual approximation.}
Let $e_a=\hat y_a-y$ for arm $a\in\{I,N\}$. Estimate its two-dimensional mean $\hat\mu_a$ and covariance $\hat\Sigma_a$ using divisor $B$. This gives five parameters per arm. With
\[
\mathcal L(\mu,\Sigma)=\E_{\xi\sim\mathcal N(0,I_2)}\|\mu+\Sigma^{1/2}\xi\|,
\]
the prediction is
\[
\widehat G_{\rm residual}=\operatorname{clip}_{[-1,1]}\left(
\frac{\mathcal L(\hat\mu_N,\hat\Sigma_N)-\mathcal L(\hat\mu_I,\hat\Sigma_I)}{\sqrt2}\right).
\]
The expectation uses fixed $40\times40$ Gauss--Hermite quadrature, with an $80\times80$ sensitivity calculation. Negative numerical covariance eigenvalues are truncated at zero. There is no learned gain mapping; the Gaussian approximation to the residual distribution is being tested.

\paragraph{Centered classification diagnostic.}
This separate diagnostic concerns the centered elliptical sign formula, not the Gaussian residual norm approximation. For $u_j=(\cos(j\pi/4),\sin(j\pi/4))$, $j=0,1,2,3$, define task margin $X_j=u_j^\top(y-\tfrac12\mathbf1)$ and output margin $M_{aj}=u_j^\top(\hat y_a-\tfrac12\mathbf1)$ for arm $a\in\{I,N\}$. On 2048 evaluation episodes per condition, a two-sided exact binomial test checks $\Prob(M_{aj}\geq0)=1/2$. Holm correction at familywise level $0.05$ covers all $20\times4\times2=160$ tests. Sign balance is necessary for centered continuous elliptical margins; nonrejection does not establish their joint ellipticity. Rejection does not rule out a noncentered Gaussian residual model, whose mean is estimated rather than fixed at zero.

\subsection{Identifiability of moment descriptions}
Let $U,S$ be independent uniform signs, let $W$ independently take the values $\pm1/2$ with equal probability, and define $X=US$. The receiver observes $(U,W)$ and the sender observes $V=SA$, with $A$ independent. In world $+$,
\[
\Prob(A=1/4)=4/5,\qquad \Prob(A=-1)=1/5;
\]
world $-$ reverses the sign of $A$. Communication produces margin $I=UV=XA$, whereas the null uses $N=UW$. Both use the same multiplication rule and sign decision.

\begin{proposition}
For $Z=(X,I,N,U,V,W)$, all joint raw moments of total degree at most three match between the two worlds. Their fixed-readout gains are $+0.3$ and $-0.3$. Any predictor restricted to those population moments and the common structural information has worst-case absolute error at least $0.3$.
\end{proposition}
\begin{proof}
Both amplitude laws have $\E A=0$ and $\E A^2=1/4$. In a monomial of degree at most three, the exponent of $A$ is the combined exponent of $I,V$. Exponents zero, one, and two depend only on identical quantities. Exponent three requires three $I/V$ factors and hence an odd independent sign $S$, giving expectation zero. The receiver's local observations and the sender's observation each leave $X$ uniform, so each local Bayes accuracy is $1/2$. The interacting fixed decoder is correct exactly when $A>0$, giving accuracies $4/5$ and $1/5$. A common prediction cannot be closer than half their gain separation to both.
\end{proof}
All variables are bounded by one and relevant margins are separated from zero. The complete unlabeled observation law of $(U,V,W)$ also matches; the labeled joint laws do not. 

The endpoints alone do not establish a difference in optimal collective capability: their amplitudes allow a distribution-aware decoder to recover the task perfectly. Mixing the two amplitude laws with weight $\theta$ instead gives fixed gain $0.6\theta-0.3$ and Bayes gain $|\theta-1/2|$. The mixtures $\theta=1/2$ and $1$ retain the same specified moments and unlabeled observations, but Bayes gains are zero and $1/2$. The corresponding worst-case error lower bound is $1/4$. These are distinct statements about fixed and optimal decoders.

The bound concerns the specified population information. It does not cover arbitrary sample statistics whose sampling distributions can contain higher-order information. The fixed-amplitude special case does not admit this obstruction: for sign-valued $X,I$, accuracy is already $(1+\E XI)/2$. This sufficient special case limits the scope of the obstruction.

\subsection{Restricted distributional sufficiency}
When the actual margins $(X,I,N)$ have a jointly centered elliptical distribution, with nonzero finite variances and no mass on decision boundaries, the established quadrant identity \cite{schmid2006} gives
\begin{equation}
G=\frac{\arcsin\rho_{XI}-\arcsin\rho_{XN}}{\pi}.
\end{equation}
Positive radial scaling preserves signs, explaining why the formula extends beyond Gaussian distributions. The assumption concerns the actual decision margins, not merely Gaussian inputs before nonlinear transformations. For nodewise scores, one averages nodewise expressions; generally two network-averaged correlations are insufficient.
The evaluated Pearson plug-in substitutes sample correlations for the population correlations, clipping them to $[-1,1]$. It does not fit an additional response curve. Population sufficiency therefore leaves a separate finite-sample estimation question.

\subsection{Dimensionless tracking predictors}
Order the Laplacian eigenvalues as $0=\lambda_1<\lambda_2\leq\cdots\leq\lambda_n$. For the tracking model, using dimensionless time $u=at$ yields
\[
\Pi=\frac{c\lambda_2}{\omega},\quad R=\frac{\omega}{a},\quad
\eta=v,\quad d_k^{(2)}=\lambda_k/\lambda_2,
\]
with modal response $H_k=1+\Pi R d_k^{(2)}$. Stationary second-moment equations give
\begin{align}
E_k(H)&=s_k\left[1+\frac{1-2H}{H(H+R)}\right]+\frac{\eta}{H},\nonumber\\
G&=\frac1n\sum_k\{E_k(1)-E_k(H_k)\}.
\end{align}
The ratio $\Pi$ cannot eliminate task spectrum $s_k$, graph spectrum $d_k^{(2)}$, response ratio $R$, or noise strength. In particular, the communication relaxation time is not an end-to-end task completion time.
\paragraph{Reference and compact predictors.}
The ratio predictor evaluates the exact formula on an eight-node ring with $R=1$ and $\eta=0.2$, varying only $\Pi$. Common and white tasks use their own reference spectra; Fiedler and high-frequency tasks use the white-task reference. The predictor therefore receives task-class information in addition to $\Pi$.

The compact predictor receives $(\Pi,R,\eta,n,s_1,\bar d)$, where $\bar d=(n-1)^{-1}\sum_{k=2}^n d_k^{(2)}$. It retains the zero mode and replaces every nonzero mode by
\[
\tilde s_k=(n-s_1)/(n-1),\qquad
\tilde H_k=1+\Pi R\bar d\quad(k>1).
\]
Substituting $(\tilde s_k,\tilde H_k)$ in the exact modal sum defines its prediction. The full-spectrum reference keeps all $(s_k,d_k^{(2)})$; its dimension grows with $n$. These mechanism-based predictors know generating parameters and are not equal-information-budget alternatives to sample-only direct calibration.

\subsection{Conditional similarity and gain reversal}
For the conditional similarity analysis, replace spatially white observation noise by noise aligned with the task spectrum. Let $p_k=s_k/n$, $\lambda_*=\sum p_k\lambda_k>0$, $\delta_k=\lambda_k/\lambda_*$, and $b=c\lambda_*/a$. All task modes share OU rate $R$. In dimensionless time $u=at$, each modal estimator obeys
\[
 dz_k=[x_k-(1+b\delta_k)z_k]du+\sqrt{2\eta s_k}\,dV_k.
\]
Thus the modal noise intensity is exactly $2\eta s_k$, with node-averaged intensity $2\eta$ because $\sum_k s_k=n$. The parameter $\eta$ is half this average intensity. The null sets $b=0$ with the same input processes. This differs from the white-noise model above. Here $\delta_k=\lambda_k/\lambda_*$ replaces $d_k^{(2)}=\lambda_k/\lambda_2$; the two normalizations are not interchangeable. With $t=b/(1+b)$ define
\begin{equation}
F(R,\eta,b)=\eta t-\frac1{1+R}+\frac{1-t^2}{1+R-Rt}.
\end{equation}
If $v_d=\sum p_k(\delta_k-1)^2$, Taylor expansion of the exact modal sum cancels the first-order term and gives
\begin{equation}
|G-F|\leq (3+2R+\eta)b^2v_d.
\end{equation}
Appendix~\ref{app:bandwidth} derives this global bound and the tighter support-dependent bound used in the experiments. For two systems satisfying the stated assumptions and matching $(R,\eta,b)$, their gains differ by at most the sum of their respective error bounds. This establishes conditional similarity, not universal curve collapse.

The nonzero root of the predicted curve, when $\eta(1+R)<1$, is
\[
b_* =\frac{R+\eta(1+R)^2}{1-\eta(1+R)}.
\]
Otherwise there is no finite positive root. This is a reversal caused by excessive smoothing, not the onset of intelligence: predicted gain is already positive at small positive coupling.

\subsection{Delay and internal stability}
The delayed controller applies delay to the full relative-state term $Lz(t-\tau)$, including cached self-state. For $\theta=a\tau$ and modal coupling $q=b\delta_k\geq0$, each mode has characteristic equation $s+1+q e^{-s\theta}=0$. If $q>1$, its critical delay is
\[
\theta_c(q)=\frac{\arccos(-1/q)}{\sqrt{q^2-1}};
\]
for $q\leq1$ every finite delay is stable. Internal network stability depends on $q_{\max}=b\lambda_{\max}/\lambda_*$.

\subsection{Applicability checks for native systems}\label{sec:gates}
Direct use of the tracking similarity law requires all of the following: a mapping to the declared scalar linear controller and its time parameters; the same nodewise squared-error task; a nonzero task spectral mean $\lambda_*$; the specified task--noise spectral alignment; and the assumed temporal covariance structure. A failed structural requirement results in an undefined prediction (NA), not a fitted surrogate parameter. No gain threshold is used to decide applicability.

The checks use the environment definition, architecture, and calibration inputs before examining new evaluation gains. Spectral summaries of the input field are diagnostics only, not replacements for the localization label. A scalar parameter is not inferred from a nonlinear Jacobian merely to make the mapping possible. The test concerns this direct transfer of the law, not every possible future effective representation.

\section{Experiments}\label{sec:experiments}
The evaluation asks three questions: can the descriptors distinguish different gains, under what conditions does a compact law hold, and is the description applicable and useful at a finite calibration budget? Each question uses the model suited to it. Gains and RMSEs are compared within tasks, not pooled across their different loss functions.

\paragraph{Common evaluation protocol.}
Predictions are evaluated on held-out conditions under the archived study-specific freezes. ``Frozen'' refers to those local protocols, not a single preregistration of the entire research sequence. The sequence was adaptive; its chronology is disclosed in Appendix~\ref{app:chronology}. Direct calibration estimates the paired mean of $Q_I-Q_N$ on target-condition samples. Where sample budgets match, descriptors and direct measurement receive the same calibration examples; extra offline training is disclosed. Known-parameter spectral predictions are not described as equal-information-cost competitors to sample-only direct measurement.

Historical analyses used positive-gain thresholds such as $0.02$. We retain their records but center the present argument on continuous gain errors, exact separations, and assumption-dependent bounds. No universal numerical onset threshold is inferred.

\subsection{Q1: Can the descriptors distinguish interaction gains?}\label{sec:q1}
This question separates empirical transfer failure from exact nonidentifiability. The transfer tests evaluate fixed prediction rules; the bounded and spectral constructions isolate pairs that share specified descriptor values.

\paragraph{Frozen transfer across distributions and tasks.}
The factorized transfer test uses 432 Gaussian training runs and evaluates 1440 runs across Gaussian, binary-latent, binary-observation, and channel-separated nonlinear families. Structural descriptors were insensitive to experimentally varied task alignment. Nevertheless, the frozen co-reachability predictor retained RMSE $0.00352$ on the aligned Gaussian subset, compared with $0.10760$ for binary latent variables and $0.09443$ for binary observations. These results confirm the original predictor's within-family accuracy while exposing its distributional limits.

Task-supervised channel reliabilities improved prediction. However, independent channels and a parity readout imply the elementary identity $G=\tfrac12\operatorname{mean}_i(r_{ai}r_{bi})$. Its success is explained by the factorized task, not arbitrary cross-task universality. The initial equal-budget direct comparison was added post hoc.

The nonseparable-task test trains the margin3 predictor defined in Section~\ref{sec:predictors} on linear-sum tasks and evaluates previously unseen radial and mixed tasks. At 512 calibration samples, RMSE of four-repeat condition means was $0.06238$, versus $0.00600$ for direct measurement. For independent Gaussian inputs alone the corresponding errors were $0.02575$ and $0.00638$, so the failure did not depend entirely on heavy tails. Single-run aggregate errors were $0.06558$ and $0.01316$: the condition-mean errors therefore do not represent single-run deployment precision. The descriptor additionally consumed offline training data. These experiments reject this predictor's transfer claim, not all low-dimensional representations.

\paragraph{Exact collision and numerical verification.}
Rational arithmetic checks all 84 monomials including the constant. Across six mixture weights and 32 simulation repeats, the maximum absolute deviation of the mean gain from its exact value is approximately $0.00100$.

\begin{figure*}[t]
\centering\includegraphics[width=0.88\textwidth]{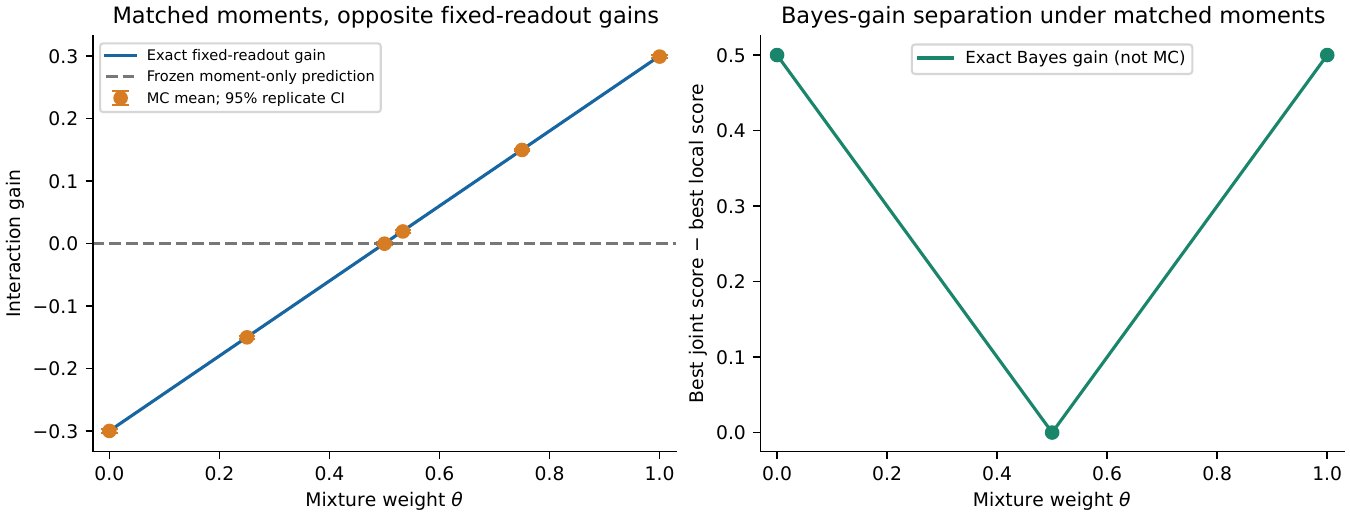}
\caption{Bounded moment-matching construction. Fixed-readout gain and distribution-aware optimal gain are different quantities. Matching population moments obstructs prediction even when fitting capacity is unrestricted. Numerical sampling checks the construction rather than proving the lower bound. The curves are analytical:
$G_{\mathrm{fixed}}=0.6\theta-0.3$ and
$G_{\mathrm{Bayes}}=|\theta-1/2|$.}
\label{fig:matched-moments}
\end{figure*}

Figure~\ref{fig:matched-moments} contrasts fixed-readout and optimal gains. The sampled curves check implementation; the lower bounds follow from the exact construction in Methods.

\paragraph{Matched spectral descriptors.}
The tracking test evaluates 585 conditions with four seeds, across sizes 8, 16, and 32, ring/path/star/complete graphs, task speeds, and coupling levels. Ten reference conditions defined a frozen ratio curve; 575 were held out. Held-out RMSE was $0.149213$ for the ratio curve, $0.118731$ for a six-scalar approximation, and $0.019663$ for the 256-sample direct reference. The full-spectrum formula has zero error against its own analytic target, which is not independent validation; comparison against independent covariance sampling gives RMSE $0.002354$.

A sharper obstruction holds on a 16-node ring. Two tasks match $(\Pi,R,\eta,n,s_1,\bar d)$ and every node's temporal target marginal, but place task power in low versus high spatial modes. Exact gains are $+0.180123$ and $-0.225757$, imposing worst-case error at least $0.202940$ for those descriptors. An additional no-sharing null retaining local degree-induced decay gives gains $+0.288556$ and $-0.117324$. Thus the sign separation is not explained solely by the change in local decay. It remains specific to a fixed smoothing strategy; optimal communication could ignore unhelpful messages.

\subsection{Q2: Under what conditions does a compact law hold?}\label{sec:q2}
The positive reference classes test whether structural assumptions can restore prediction, while controls identify the information those assumptions supply. Numerical agreement within a constructed class is not evidence of unconditional universality.

\paragraph{Elliptical reference design.}
We use $\rho_{XI}\in\{-0.6,0,0.6\}$ and $\rho_{XN}\in\{-0.3,0,0.3\}$, with eight repeats per pair and family. Starting with independent standard Gaussian coordinates $g_0,g_1,g_2$, margins are $X=g_0$, $I=\rho_{XI}g_0+\sqrt{1-\rho_{XI}^2}g_1$, and analogously $N$ using $g_2$. The bounded family first scales $g$ by $\sqrt3/\|g\|$; the common-radial $t_3$ family scales it by an independent $1/\sqrt{\chi_3^2}$. These preserve the sign relation. Each of the 216 runs across the three families uses 512 calibration observations and 32768 separate evaluation observations; a 64-example prefix provides a prespecified sensitivity check at a smaller budget. The exact arcsin gain is the reference for estimation RMSE. Finite-budget results are reported under Q3.

\paragraph{Matched tracking systems.}
The conditional-law test uses second-neighbor circulant rings, comb graphs, and three-clique modular graphs with bridge weight $0.15$, at sizes 18, 36, and 72. Two task-band locations per graph/size select the three nearest positive modes around the 0.25 and 0.65 quantile indices, with equal power, giving 18 systems. We fix $R\in\{0.3,2\}$, $\eta\in\{0.1,0.5\}$ and $b\in\{0.05,0.1,0.2,0.4,0.8,1.6,3.2,6.4\}$. Each task spectrum remains fixed along its coupling curve.

The narrow-band condition aligns modal noise with task power. A broad-spectrum control places power at zero and maximum eigenvalues while preserving $\lambda_*$ and total power; a white-noise control preserves the task and total noise power but removes spectral alignment. Matching $(R,\eta,b)$ therefore isolates the assumptions omitted by those controls. No empirical precision threshold defines curve collapse: we report prediction error, within-group spans, and the analytical error bound.

\begin{figure*}[t]
    \centering
    \includegraphics[width=\linewidth]{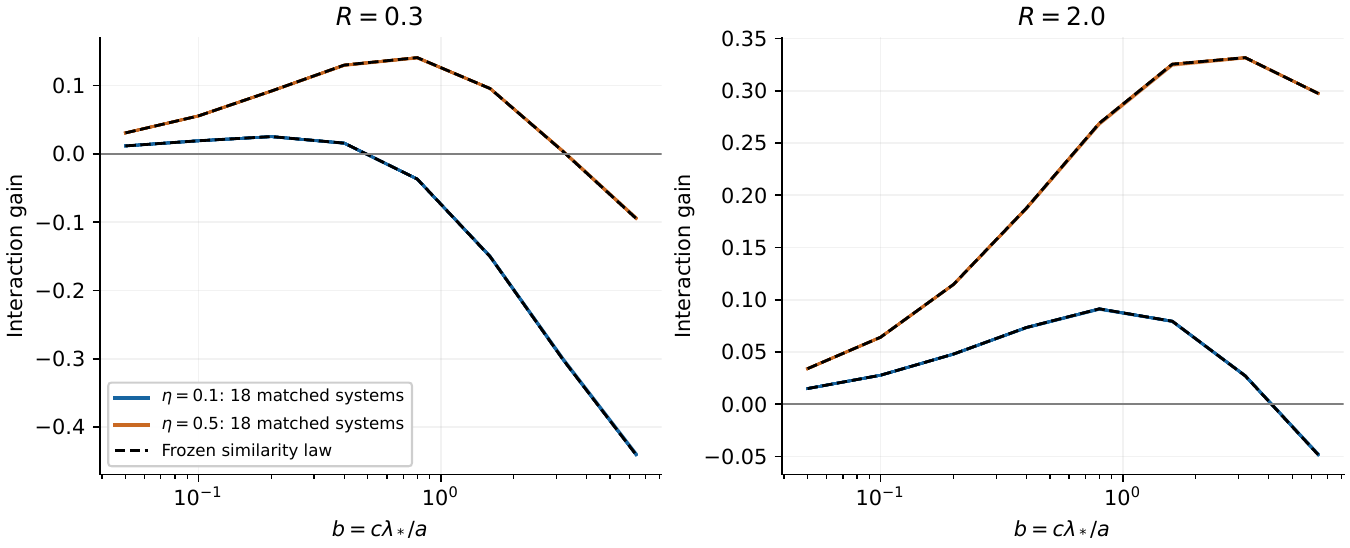}
    \caption{Conditional similarity across matched tracking systems.
    Panels fix $R$; colors distinguish $\eta$.
    For each $(R,\eta)$ pair, solid curves show gains for
    18 constructed systems, and black dashed curves show
    the frozen analytical prediction.
    The horizontal axis is $b=c\lambda_*/a$.
    Systems satisfy narrow task-band and noise-alignment
    assumptions; agreement supports similarity within this class.
    The broad-spectrum and white-noise controls reported
    in the text delimit its applicability.}
    
\label{fig:conditional-collapse}
\end{figure*}

\paragraph{Curve-collapse results.}
Figure~\ref{fig:conditional-collapse} shows the matched gain curves. Across 18 constructed systems from three graph families and sizes 18/36/72, the 576 narrow-band aligned zero-delay conditions yield RMSE $0.0002791$ and maximum error $0.002132$. Seven systems have degenerate equal-eigenvalue task bands; on the remaining 11 systems RMSE is $0.0003570$. Task rank remains three as size changes.

Across the 32 zero-delay narrow-band groups matching $(R,\eta,b)$, each containing 18 systems, the median gain span is $0.000419496$ and the maximum is $0.002131674$. The archived bound checks report no violations in either the 576 narrow-band or the 576 broad-spectrum aligned zero-delay conditions, allowing for recorded numerical error. These checks use the support-dependent bound $B_0$ in Eq.~\eqref{eq:support-bound}, rather than only the global bound displayed in the main text. Its median is $0.0000221453$ in the narrow-band class and $0.351508$ in the broad-spectrum control: validity alone does not ensure a sharp prediction. The exact records are \path{studies/round10/results/collapse_spans.csv} and \path{studies/round10/results/metrics.csv}; Appendix~\ref{app:bandwidth} gives the support-dependent derivation used by these records. Broad-spectrum and spatially white-noise controls yield RMSE $0.073970$ and $0.029597$, respectively, and retain sign disagreements. The broad-spectrum control weakens the narrow-band approximation but remains within the aligned-noise bandwidth bound; the white-noise control violates the alignment assumption, so that bound does not apply unchanged.

\paragraph{Delay stress test.}
The delay grid is $\theta\in\{0,0.02,0.1,0.3,0.8\}$, applied to all three spectral mechanisms, for 8640 numerical records. Stability is checked before stationary gain is computed. Zero-delay gains are checked against modal Lyapunov equations; stable delayed gains use the archived frequency-domain calculation. Systems beyond the exact stability boundary remain marked unstable. The frozen grid contains 32 groups sharing $(R,\eta,b,\theta)$ with both stable and unstable members. Task-response similarity therefore cannot replace an all-mode stability check. The auxiliary delay-corrected predictor and its empirical scope are defined in Appendix~\ref{app:delay}; they are not used as a proved finite-bandwidth similarity result. Unstable cases receive no fictitious stationary gain.

\subsection{Q3: Is the description applicable and useful at a finite calibration budget?}\label{sec:q3}
This question tests estimation efficiency and applicability separately. A known population formula may be statistically inefficient to estimate, and a valid formula for one controller need not define a prediction for another.

\paragraph{Population sufficiency versus estimation efficiency.}
With 512 calibration observations, the experiment gives the following single-run RMSEs against exact gain:
\begin{center}
\begin{tabular}{lrr}\toprule
Distribution & Pearson formula & Direct\\\midrule
Gaussian & .01819 & .03204\\
Bounded elliptical & .01466 & .03204\\
Common-radial $t_3$ & .05903 & .03204\\\bottomrule
\end{tabular}
\end{center}
Thus a known sufficient population description can help in some distributions and lose in another. The $t_3$ result concerns this estimator, not all robust estimators. The three family experiments share sampled directions, so they are not three independent replications. The classification transfer study also included squared heavy-tailed margins lacking finite second moments; those are explicitly outside the second-moment assumptions.

\begin{figure*}[t]
\centering\includegraphics[width=0.88\textwidth]{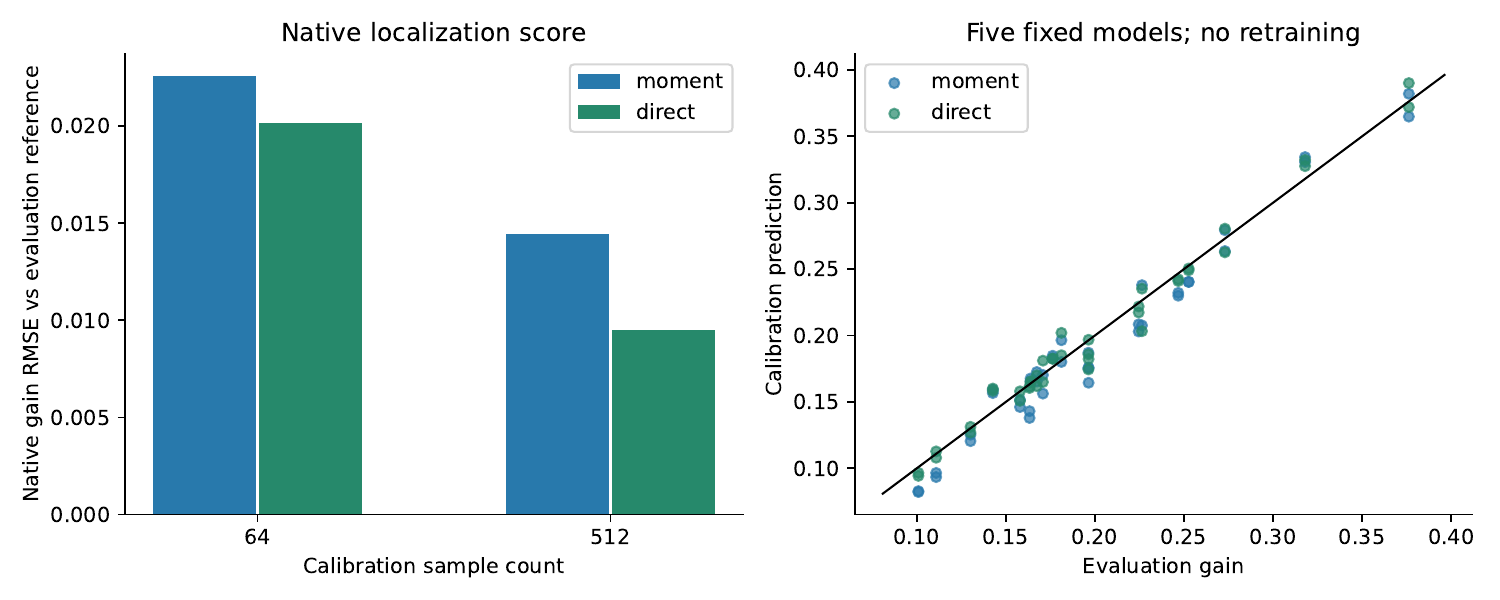}
\caption{Native localization prediction at matched calibration budgets. The Gaussian moment approximation does not outperform direct measurement at matched target-condition calibration budgets. The null retains the same frozen weights and global mean readout.}\label{fig:native-prediction}
\end{figure*}

\paragraph{Native cellular prediction.}
Figure~\ref{fig:native-prediction} summarizes the native prediction comparison. The cellular comparison uses five fixed models and four sensor-density/noise conditions, totaling 20 conditions. Each condition has 1024 calibration episodes split into two 512-example blocks and 2048 independent evaluation episodes. A fixed Gaussian residual approximation uses ten mean/covariance parameters for the two-dimensional interaction and null residuals; it is not a direct application of the zero-threshold elliptical sign formula.

At budget 512, its RMSE is $0.014434$, versus $0.009498$ for the paired direct estimate, a 51.97\% increase. At budget 64 the errors are $0.022581$ and $0.020149$. The model-cluster bootstrap interval for the 512-budget RMSE difference is $[0.000374,0.010938]$, conditional on only five previously examined checkpoints. Evaluation gain is itself a finite-sample reference, with mean standard error $0.003258$. Thus these are not errors against known population truth or reliable confirmation over a new population of trained models.

All 20 estimated gains in this experiment are positive, ranging from $0.100859$ to $0.376212$. This supports usefulness of the declared coupling intervention in these systems, not a capability-onset boundary or superiority over an optimized independent architecture. The separate centered-classification diagnostic defined in Section~\ref{sec:predictors} rejects sign balance in 132 of 160 tests after Holm correction. This limits direct use of the centered sign formula; it does not establish why the noncentered Gaussian residual predictor underperforms.

\paragraph{Applicability to the unchanged cellular system.}
The applicability test applies Section~\ref{sec:gates} to the same five checkpoints and four environments: sensor density $0.05/0.2$ crossed with noise standard deviation $0.1/0.3$. Each condition has 128 calibration and 128 independent evaluation episodes; no parameter mapping is refitted. The architecture and generator already violate the required controller, loss, and temporal correspondences. All 20 conditions are rejected, so no tracking-law prediction is issued and transfer RMSE is undefined. Static sensor noise reused through internal iterations is not a temporally white observation process, global localization loss is not nodewise tracking loss, and coordinate-dependent nonlinear responses do not share uniform scalar modal dynamics. Input-field spectral diagnostics are not substituted for the actual localization task.

\paragraph{Task relevance within the native system.}
As a supporting diagnostic of task relevance, we divide the lattice into four $16\times16$ regions and use 64 evaluation episodes per condition. Three regions receive episode observations shifted by independently chosen nonzero cyclic offsets; the fourth keeps its original evidence. This preserves each region's empirical input distribution while changing its relation to the target. The paired coherent control uses the same masks and reconstructed sensor-noise residuals as the scrambled inputs, but restores the clean field for the correct target. The native task, weights, and communication/null comparison remain fixed. Mean gain is $0.209373$ in the paired coherent control and $0.118317$ after scrambling; all 20 differences are positive, but all scrambled-condition gains remain positive too. The manipulation changes task-observation relationships and inference difficulty together. Reused episodes and shared checkpoints preclude treating these as 20 independent confirmations of statistical significance. This is evidence that task consistency matters under this intervention, not identification of a unique synergy mechanism or an intrinsic integration timescale.

\section{Discussion}\label{sec:discussion}

\subsection{What the evidence establishes}
The results separate three questions often conflated in the search for a collective control parameter. A descriptor can fail because distinct systems share its value, because an assumed distributional or dynamical class does not apply, or because estimating it is less effective than measuring gain directly. Conversely, a population identity or a conditional collapse can be valid without yielding a practically useful predictor at the evaluated calibration budgets.

The three model classes address different parts of this separation. The bounded construction supplies exact information restrictions, the tracker supplies an analyzable mechanism, and the cellular system supplies a nonlinear applicability test. Their juxtaposition does not establish a common microscopic mechanism. Their relevance to embodied biological collectives remains a question for models with appropriate sensing, action, and environmental feedback.

The evidence is strongest for specified policies and comparison interfaces. Negative gain under a fixed policy does not mean that additional information lowers Bayes-optimal performance. Likewise, architectural coupling interventions can change computations as well as information access. These qualifications determine the scientific meaning of the measured contrast.

\subsection{Effective agents as a future direction}
A future route is to seek a state or spatial coarse-graining $C$ such that the projected complex dynamics and effective dynamics approximately agree under corresponding inputs and interventions. For stochastic transition kernels, the target relation is
\[
C_{\#}P_{\alpha}^{(k)}(x,\cdot)\approx
\bar P_{\bar\alpha}(C(x),\cdot).
\]
The same map should work for interaction and its null, with explicit observation permissions, readout, time correspondence, and acquisition cost. If hidden microstates materially affect the next coarse state, memory may be required. A local linearization alone neither reduces dimension nor proves this relation.

For the paired interaction/null comparison \cite{wu2026}, if armwise expected quality errors are bounded by $\epsilon_I,\epsilon_N$, then $|G-\bar G|\leq\epsilon_I+\epsilon_N$. This triangle-inequality consequence is not a new certificate. Establishing its premises from a useful coarse model is the unsolved problem. Classification additionally requires control near decision boundaries; multi-step approximation requires error-propagation control. Wrong certified signs, unpreserved nulls, omitted unstable modes, and excessive parameter-acquisition cost are explicit failure outcomes.

The scientific ambition is that different micro-level implementations may share effective task-relevant dynamics and hence a small set of dimensionless parameters. Neither that effective equivalence nor a reduction among the three present models has been demonstrated. This direction should be tested on the existing systems without altering them to enforce the desired theoretical assumptions.

\subsection{Limitations and reproducibility}
The research sequence is adaptive, while individual studies have archived freezes. Known generating parameters underpin the strongest spectral results. The conditional-collapse systems were constructed around their hypotheses; they are not a blind test on naturally occurring populations. Neural results reuse five trained checkpoints and a narrow localization task. There is no universal threshold, finite-size phase-transition evidence, or demonstrated low-cost acquisition of sufficient spectral parameters.

Code, configuration freezes, raw outputs, and study-specific reproduction records are organized in the prepared archive \path{RCI_Paper2_reproducibility.zip}; Appendix~\ref{app:evidence} maps results to its relative paths. As of this working-manuscript revision, the archive is held locally and no public repository or deposit URL has been established. The LaTeX source package includes the manuscript, bibliography, and existing figures, but not the full experimental archive. Public availability of the latter is therefore pending; the existence of local records should not be read as a claim that readers can already download them. Existing records report same-author fresh-process replays; these are not independent external replications. Protocol exceptions, including a cellular wall-time overrun and post-hoc comparisons, are documented in the chronology appendix. The package preserves existing outputs and identifies analyses that were post hoc; its integrity checks do not substitute for a fresh scientific replay.

\section{Conclusion}
Low-dimensional prediction of collective task gain is possible in explicit restricted classes, but neither dimensionality nor population sufficiency ensures transfer or finite-budget usefulness. Bounded descriptor collisions, task-spectrum gain reversals, and failed cellular applicability checks identify distinct obstacles. Conditional curve collapse provides a positive mechanism result without eliminating those obstacles. A collective-intelligence Reynolds number remains a scientific objective: it will require independently meaningful parameters, preserved comparison interfaces, and evidence that effective descriptions survive changes of task, scale, and implementation.

\bibliographystyle{plain}
\bibliography{references}
\clearpage
\appendix
\section{Evidence map and reproducibility}\label{app:evidence}
The prepared reproducibility archive identified in Section~\ref{sec:discussion} retains study directories under \texttt{studies/}. Its original README uses the earlier working title, \emph{Predicting Interaction Gains in Collective Systems: Conditions and Limits of Low-Dimensional Descriptions}; it contains the same selected historical studies. This archive is separate from the manuscript source ZIP, and its public deposition is pending. Directory labels record the historical sequence, not separate claims of independent confirmation. Table~\ref{tab:evidence} maps the selected results to their archive locations. Each directory contains its protocol, code, outputs, and interpretation notes. The package-level \texttt{verify\_package.py} verifies file hashes and selected table values without generating experimental observations. After extracting that archive, \texttt{python3 verify\_package.py} performs this saved-artifact check. Study-specific replay commands can generate new observations and are a different operation; they have not been executed for this manuscript revision. Historical replay records are supplied as evidence of previous same-author re-execution, not independent replication in this release.

\begin{table*}[t]
\centering\small
\begin{tabular}{p{.16\textwidth}p{.34\textwidth}p{.42\textwidth}}
\toprule
Study directory & Principal evidence & Scope of verification\\\midrule
\texttt{pilot} & Frozen factorized inference transfer & \texttt{RESEARCH\_MEMO\_ZH.md}; \texttt{results/}; original coefficient provenance\\
\texttt{round2} & Nonseparable-task prediction vs. direct calibration & \texttt{results/metrics.csv}; distinguish run and condition summaries\\
\texttt{round3} & Bounded moment collision and Bayes distinction & \texttt{PROOF\_ZH.md}; \texttt{results/summary.json}; rational moment checks\\
\texttt{round4} & Elliptical sufficiency and estimation cost & \texttt{THEORY\_ZH.md}; \texttt{results/metrics.csv}\\
\texttt{round5} & Native cellular prediction & \texttt{results/metrics.csv}; \texttt{predictions.npz}; fixed weights and native readout\\
\texttt{round6} & Dimensionless tracking and spectral collision & \texttt{DERIVATION\_ZH.md}; \path{results/prediction_metrics.csv}\\
\texttt{round10} & Conditional similarity and delay boundary & \texttt{THEORY\_ZH.md}; \texttt{results/metrics.csv}; changed noise model\\
\texttt{round11} & Applicability rejection on native systems & Frozen gates, raw diagnostics, and \texttt{RESEARCH\_MEMO\_ZH.md}\\
\texttt{round14} & Task-consistency intervention & Frozen intervention and \texttt{RESEARCH\_MEMO\_ZH.md}; shared-model limitations\\
\bottomrule
\end{tabular}
\caption{Evidence locations relative to \texttt{studies/}. Original study files and negative conditions are retained.}\label{tab:evidence}
\end{table*}

\subsection{Design details for the transfer studies}\label{app:transfer}
\paragraph{Factorized transfer model.}
Let $W$ be the row-normalized neighbor matrix, $P=(1-\alpha)I+\alpha W$, $K=\gamma^T P^T$, and $K_0=\gamma^T I$. The role vector $b$ assigns each node to one of two channels. Latents $(z_1,z_2)$ are independent standard Gaussians in the Gaussian and nonlinear families, and independent uniform signs in the two binary-latent families. The second observed latent is $\tilde z_2=z_2$, $-z_2$, or an independent draw from the same marginal, representing aligned, reversed, or unrelated evidence. Except in the binary-observation family, node inputs are $A_i=(1-b_i)(z_1+\epsilon_i)$ and $B_i=b_i(\tilde z_2+\epsilon_i)$, with independent standard Gaussian sensor noises across nodes and episodes. In the binary-observation family, replace the parenthesized terms by $\sqrt2 z_1 f_i$ and $\sqrt2\tilde z_2 f_i$, where $f_i=-1$ with probability $(1-1/\sqrt2)/2$ and $+1$ otherwise. The two-channel readout noise has independent Gaussian coordinates with standard deviation $\sigma$.

Linear-family outputs are $(KA,KB)$ plus readout noise; the null substitutes $K_0$. Both arms share sensor and readout draws. The nonlinear family instead iterates $v^{t+1}=\tanh(\gamma Pv^t)$ separately on each channel, with $P=I$ in the null, adding readout noise only at the end. The terminal decision is the sign of the product of the two output channels; the target is $\operatorname{sign}(z_1z_2)$. This factorized study uses the conventional sign function, including $\operatorname{sign}(0)=0$, and does not divide outputs by channel response mass.

\paragraph{Nonseparable transfer model.}
This study uses independent standard Gaussian latents, correlated Gaussian latents with standard normal marginals and correlation $0.7$, or independent $t_3/\sqrt3$ latents. Aligned, reversed, and unrelated second-channel observations are generated as above, with an independent same-family marginal for unrelated evidence. Sensor and readout noises remain Gaussian. Let $a_1=1-b$, $a_2=b$, $\tilde z_1=z_1$, and define for arm matrix $K_a\in\{K,K_0\}$
\[
 O^{(a)}_{ij}=\frac{[K_a(a_j\odot(\tilde z_j\mathbf1+\epsilon))]_i+\xi_{ij}}
 {[K_a a_j]_i},\qquad j=1,2.
\]
The implementation assigns $O^{(a)}_{ij}=0$ when the denominator is at most $10^{-12}$. Here $\epsilon_i\sim\mathcal N(0,1)$ and $\xi_{ij}\sim\mathcal N(0,\sigma^2)$ are independent sensor and readout noises. The paired arms share $\epsilon$ and $\xi$, but use their own known response masses. This normalization uses model information, not task labels or pooled observations.

The four task margins are $m_{\rm parity}(z)=z_1z_2$, $m_{\rm sum}(z)=(z_1+z_2)/\sqrt2$, $m_{\rm radial}(z)=z_1^2+z_2^2-2\log2$, and $m_{\rm mixed}(z)=z_1+z_2+0.8z_1z_2$. The task label is $\mathbf1[m(z)\geq0]$ and each node uses $\mathbf1[m(O_i^{(a)})\geq0]$; zero margins are classified positive. Episode quality averages node correctness. Parity is the factorized control, sum is the training task, and radial/mixed are the primary unseen tasks. These decision conventions apply to this study, not retroactively to the factorized sign-based study.

The factorized study used 512 independent calibration tasks and 1024 evaluation tasks per test run. Its 1440 test runs combine five held-out condition categories, eight configurations per category, three alignments, three seeds, and four families. Reported condition-level metrics average three seeds. Task-supervised descriptors observe component labels; structural descriptors do not. The corresponding information difference must accompany any performance comparison.

The nonseparable study trained on 192 Gaussian linear-sum runs and tested 864 base runs, each evaluated on four tasks at two calibration budgets, yielding 6912 task--budget records. These records share observations within each base run and are not independent replications. Each run uses 512 calibration examples (64-example prefixes form the smaller-budget analysis) and 4096 independent evaluation examples. Interaction and null share a parameterized decoding rule but use their respective known channel-response normalizations. This is not a shared set of numerical decoder coefficients; the normalizations use graph/model information, not pooled observations. Bootstrap intervals condition on the finite design and do not certify generalization to arbitrary tasks.

\subsection{Frozen co-reachability coefficients}\label{app:coefficients}
The original co-reachability predictor uses the ordered feature vector
\[
\psi=(C,C^2,C^3,CH_c,CB_c,CH_cB_c)^\top
\]
and $\widehat G_{\rm coreach}=\operatorname{clip}_{[0,1/2]}(\beta_{\rm coreach}^\top\psi)$, with no intercept. The frozen coefficient vector is
\[
\beta_{\rm coreach}=\begin{pmatrix}
0.020771702942835257 \\
0.4552567620070011 \\
0.11121442609127323 \\
-0.08983172717917717 \\
0.28040798228192665 \\
0.7174118023920321
\end{pmatrix}.
\]
These values are copied from the original \texttt{coreach\_three} entry in \path{pilot/vendor/probe2_frozen.json}, relative to the archive's \texttt{studies/} directory. The original ridge penalty was $10^{-6}$. No coefficients were refitted for the present comparisons. This predictor is distinct from the task-reliability and Margin3 fits, which use ridge penalty $0.01$.

\subsection{Exact Bayes calculation for the mixture}
At each observable amplitude, let the two component amplitude laws be mixed with weight $\theta$. Conditional on an observation, the task sign has posterior probabilities $\theta$ and $1-\theta$, possibly with their labels reversed. The optimal joint accuracy is therefore $\max(\theta,1-\theta)$, while each local accuracy stays $1/2$. This gives $G_{\rm Bayes}=|\theta-1/2|$. The unlabeled observation distribution is identical across mixture weights because the independent sign $S$ symmetrizes $A$. Consequently, labeled nonlinear information, rather than the unlabeled marginals, distinguishes the worlds. Negative fixed-decoder gain cannot occur for an unconstrained optimal decoder that can ignore messages.

\subsection{Bandwidth bound}\label{app:bandwidth}
Write $h(q)=\eta q/(1+q)+f_R(q/(1+q))$, where
\[
f_R(t)=-1/(1+R)+(1-t^2)/(1+R-Rt).
\]
For $R\geq0$ and $t=q/(1+q)\in[0,1)$, $|f'_R(t)|\leq2$ and
$f''_R(t)=-2(1+2R)/(1+R-Rt)^3$.
The chain rule with $t'=1/(1+q)^2$ and $t''=-2/(1+q)^3$ yields
$|h''(q)|\leq 2(3+2R+\eta)$.
Since $\sum_k p_k(\delta_k-1)=0$, Taylor's remainder gives
\[
\left|\sum_k p_k h(b\delta_k)-h(b)\right|
\leq(3+2R+\eta)b^2\sum_kp_k(\delta_k-1)^2.
\]
This establishes the stated zero-delay bound. It assumes the aligned-noise model and does not apply unchanged to the spatially white-noise controls. The method is a standard Taylor argument, specialized here to the signed task-gain response.

\paragraph{Support-dependent bound used in the experiments.}
Let $d_\ell=\min_{k:p_k>0}\delta_k$ and $d_u=\max_{k:p_k>0}\delta_k$, and define $q_\ell=bd_\ell$, $q_u=bd_u$, and $t_u=q_u/(1+q_u)$. Since $\sum_kp_k\delta_k=1$, the expansion point $q=b$ lies in this support interval. On that interval, the same derivative calculation gives
\begin{align}
\frac12|h''(q)|&\leq C_s,\nonumber\\
C_s&=\frac{1+2R}{(1+R-Rt_u)^3(1+q_\ell)^4}\nonumber\\
&\quad+\frac{\eta+2}{(1+q_\ell)^3}.
\end{align}
Combining this estimate with $C_0=3+2R+\eta$ yields the bound used in the archived zero-delay checks:
\begin{equation}
|G-F|\leq B_0,\qquad
B_0=\min\{C_0,C_s\}\,b^2v_d.
\label{eq:support-bound}
\end{equation}
Indeed, $|f_R''(t)|$ is largest at $t_u$, while $t'$ and $|t''|$ are largest at $q_\ell$; using $|f'_R(t)|\leq2$ bounds the remaining chain-rule term. Taylor expansion then cancels the weighted first-order term as above. The support endpoints include only modes with positive task power. This support-dependent result still assumes aligned noise and zero delay; it is not a bound for the spatially white-noise controls or the delayed predictor.

\subsection{Auxiliary delayed-response predictor}\label{app:delay}
For a stable mode define $H_\theta(i\nu;q)=[1+i\nu+q e^{-i\nu\theta}]^{-1}$, $K_\theta(q)=[1+R+q e^{-R\theta}]^{-1}$, and
\[
J_\theta(q)=\frac{2R}{\pi}\int_0^\infty\frac{|H_\theta(i\nu;q)|^2}{R^2+\nu^2}\,d\nu.
\]
Let $N_\theta(q)$ be its stationary noise variance under diffusion coefficient $\sqrt{2}$ (noise intensity $2$). The auxiliary response is
\[
h_\theta(q)=2K_\theta(q)-J_\theta(q)-\frac1{1+R}+\eta[1-N_\theta(q)].
\]
For $q>1$, $N_\theta(q)=[1+q\sin(\nu_q\theta)/\nu_q]/[1+q\cos(\nu_q\theta)]$, where $\nu_q=\sqrt{q^2-1}$; the expression has the hyperbolic continuation for $q<1$ and limit $(1+\theta)/2$ for $q=1$. The predictor is $F_\theta=h_\theta(b)$ and the aligned-spectrum reference is $\sum_k p_kh_\theta(b\delta_k)$. The integral uses the archived error-controlled frequency quadrature. Predictions are evaluated only when the required modes are stable. Small errors observed on stable subsets are empirical evidence; the zero-delay bandwidth bound is not a proved uniform finite-bandwidth bound for $F_\theta$.

\subsection{Chronology and retained exceptions}\label{app:chronology}
The initial direct-calibration comparison was post hoc. The restricted common-target similarity observation in the first tracking study was also post hoc; the later narrow-band study froze its relation before its new numerical grid. The native cellular models had already been examined in earlier studies, even when new evaluation episodes were generated. The 32-step native prediction experiment is not the original 16-step pilot scan. Its original run exceeded its intended wall-time limit; recorded completion and historical replays do not erase that protocol exception. No fitting was performed to turn the failed native applicability checks into numerical predictions.
\end{document}